\documentclass[prd,english,preprintnumbers,amsmath,amssymb,nofootinbib,twocolumn,superscriptaddress]{revtex4-1}

\pdfoutput=1

\usepackage[latin1]{inputenc}
\usepackage{graphicx}
\usepackage{bbm}
\usepackage{bbold}
\usepackage{amssymb}
\usepackage{amsmath}
\usepackage{tabularx}
\usepackage{slashed}

\usepackage{dsfont}

\usepackage{ulem}
\def\0#1#2{\frac{#1}{#2}}

\def\s0#1#2{\mbox{\small{$ \frac{#1}{#2} $}}}

\newcommand{\be}{\begin{eqnarray}}
\newcommand{\ee}{\end{eqnarray}}

\newcommand{\nn}{\nonumber}

\newcommand{\beq}{\begin{equation}}
\newcommand{\eeq}{\end{equation}}
\newcommand{\bea}{\begin{eqnarray}}
\newcommand{\eea}{\end{eqnarray}}

\usepackage{babel}
\makeatother
\begin{document}

\title{On the Search for Inhomogeneous Phases in Fermionic Models}

\author{Jens Braun}
\affiliation{Institut f\"ur Kernphysik (Theoriezentrum), Technische Universit\"at Darmstadt, 
D-64289 Darmstadt, Germany}
\affiliation{ExtreMe Matter Institute EMMI, GSI, Planckstra{\ss}e 1, D-64291 Darmstadt, Germany}
\author{Stefan Finkbeiner}
\affiliation{Institut f\"ur Kernphysik (Theoriezentrum), Technische Universit\"at Darmstadt, 
D-64289 Darmstadt, Germany}
\author{Felix Karbstein}
\affiliation{Helmholtz-Institut Jena, Fr\"obelstieg 3, D-07743 Jena, Germany}
\affiliation{Theoretisch-Physikalisches Institut, Friedrich-Schiller-Universit\"at Jena, Max-Wien-Platz 1, D-07743 Jena, Germany}
\author{Dietrich Roscher}
\affiliation{Institut f\"ur Kernphysik (Theoriezentrum), Technische Universit\"at Darmstadt, 
D-64289 Darmstadt, Germany}

\begin{abstract}
We revisit the Gross-Neveu model with $N$ fermion flavors in {1+1} dimensions and compute its 
phase diagram {at finite temperature and chemical potential} in the large-$N$ limit. 
To this end, we double the number of fermion degrees of freedom in a specific way which allows us to detect inhomogeneous 
phases in an efficient manner.
We show analytically that this ``fermion doubling trick'' predicts correctly the position of the boundary between the chirally
symmetric phase and the phase with broken chiral symmetry.
Most importantly, we find that the emergence of an inhomogeneous ground state is predicted correctly. 
We critically analyze our {approach} based on this trick and discuss its applicability to other theories, such as fermionic models
in higher dimensions, where it may be used to guide the search for inhomogeneous phases.
\end{abstract}

\maketitle

%
\section{Introduction}
The search for exotic phases, such as Fulde-Ferrell-Larkin-Ovchinnikov-type phases \cite{FF,LO}, plays an important role in various
different fields of physics, ranging from condensed-matter systems (see, e.g., Refs.~\cite{Jackiw:1981wc,Mertsching,Machida,Chodos:1998wg,Kleinert:1998kj}),
over ultracold atomic gases (see, e.g., Refs.~\cite{Bulgac:2008tm,Stoof,Radzihovsky,Roscher:2013cma}), to
{exactly solvable, relativistic model field theories for the strong 
interaction (see, e.g., Refs.~\cite{Schon:2000he,Schon:2000qy,Schnetz:2004vr,Schnetz:2005ih,Basar:2008im,Basar:2008ki,Basar:2009fg}) and}
high-energy physics with an emphasis on QCD 
phenomenology (see, e.g., Refs.~\cite{Bringoltz:2006pz,Bringoltz:2009ym,Nickel:2009wj,Kojo:2009ha,Buballa:2014tba}),
{see also Refs.~\cite{Casalbuoni:2003wh,Thies:2006ti} for more general reviews.}
The latter studies 
were triggered by Thies' ground-breaking {analyses} of inhomogeneous phases in 
{Gross-Neveu type} models {in $1+1$ dimensions}~\cite{Schon:2000qy,Brzoska:2001iq,Thies:2003br,Thies:2003kk}.

Loosely speaking, the emergence of condensates in fermionic theories 
is tightly linked to the formation of bosonic two-fermion bound-states. The macroscopic
occupation of the energetically lowest lying bosonic state is then associated with the emergence of a condensate. In general, the latter goes alongside with
the spontaneous breakdown of a fundamental symmetry of the underlying fermionic theory, such as the breakdown of the U($1$) symmetry in ultracold Fermi gases indicating
the presence of a superfluid ground state.
Fulde and Ferrell as well as Larkin and Ovchinnikov suggested that the formation of a spatially varying, i.e. inhomogeneous, ground state might
be energetically favored in fermionic systems with two components, provided that the associated Fermi momenta 
are sufficiently different~\cite{FF,LO}. The latter can be controlled
by, e.g., varying the associated chemical potentials. In such a situation, 
the energetically lowest lying (bosonic) bound-state configuration may carry
a finite center-of-mass momentum. The macroscopic occupation of this state may 
then give rise to the formation of a spatially varying condensate, see, e.g., Ref.~\cite{Roscher:2013cma}
for an illustration in the context of ultracold Fermi gases.

For our understanding of the emergence of inhomogeneous phases, studies of {-- at least within the mean-field approximation -- exactly solvable} 
low-dimensional {models play} a prominent role~\cite{Schon:2000qy,Thies:2003br,Schnetz:2004vr,Schnetz:2005ih,Basar:2008im,Basar:2008ki,Basar:2009fg}.
{Noteworthily,} in the case of one-dimensional models, {the outcomes of}
mean-field studies are expected to be {of most relevance for higher dimensions:}
{Beyond {the mean-field approximation}, the presence of} long-range fluctuations hinders spontaneous symmetry breaking {in one dimension}~\cite{MW,HB}, 
rendering the results considerably distinct from higher dimensions.
{Hence, from a field-theoretical point of view, mean-field studies of a given one-dimensional model 
{are rather considered to} be useful to guide studies of the corresponding model in three dimensions where fluctuation effects often play a less prominent role.}
{In particular, the above-mentioned knowledge of analytic solutions in the one-dimensional case is very appealing since it allows 
to provide analytic guidance for, e.g., studies of QCD with imaginary chemical potential \cite{Karbstein:2006er} underlying 
many lattice {simulations (see, e.g., Refs.~\cite{deForcrand:2002ci,Lombardo:2006yc,Philipsen:2012nu}), 
and} also opens up the possibility to develop and benchmark new techniques
for, e.g., the study of the emergence of inhomogeneous phases being the subject of the present study.}
{In the latter case, the} techniques may then be applied to models in higher dimensions where the search
for the existence of inhomogeneous ground states requires in general the use of numerical {methods}.
Therefore, already one-dimensional models serve as a valuable test ground to better understand how the dynamics of a strongly coupled field theory like, e.g., QCD is
affected by the presence of a baryon~\cite{Schon:2000qy,Schnetz:2004vr,Schnetz:2005ih} 
or isospin chemical potential~\cite{Ebert:2011rg,Gubina:2012wp,Ebert:2013dda}.

In this paper, we present and discuss a ``fermion doubling trick" which allows to search for the emergence of inhomogeneous phases in
an efficient way. For illustration purposes, we use this trick to study the phase diagram of the Gross-Neveu (GN) model in one spatial dimension 
{at finite temperature and chemical potential} in the large-$N$ limit.
We show that the position of the transition line between the chirally symmetric phase and the phase with broken chiral symmetry
is recovered correctly {for this model, rendering} {the fermion doubling} trick potentially useful to 
assist the search for inhomogeneous phases in higher dimensional fermionic models.
In Sect.~\ref{sec:formalism}, we introduce our fermion doubling approach 
and {exemplarily} demonstrate the computation of the effective potential {based thereon} {for} the GN model.
Moreover, our approach and its limitations are critically {reviewed and} discussed. The phase diagram of the GN model as obtained from our doubling approach is 
presented in Sect.~\ref{sec:pd} and compared to the well-known 
exact solution~{\cite{Thies:2003kk,Schnetz:2004vr,Thies:2006ti}}. 
Our conclusions are given in Sect.~\ref{sec:conc}.

\section{Formalism}\label{sec:formalism}
\subsection{Gross-Neveu Model}
{In the present work, we refrain from}
repeating all the well-known attractive features of {the GN model but rather highlight 
the most important features relevant for our study and refer} the reader to some pertinent review articles
otherwise, see Refs.~\cite{Schon:2000qy,Feinberg:2003qz,Thies:2006ti} and the references therein.

Originally, the GN model has been introduced 
as a toy model for the strong interaction, and more specifically to study dynamical chiral symmetry 
breaking in asymptotically free fermionic theories~\cite{Gross:1974jv}.
It describes the quantum field theory
of $N$ flavors of relativistic fermions interacting via a four-fermion interaction term. Its action in {$d=$}1+1 Euclidean
space-time {dimensions} reads
\be
  S
  =\int_{\tau}\int_{x}\, \Big\{\bar\psi\left({\rm i}\slashed{\partial}+{\rm i}\mu \gamma_0\right)\psi 
  + \frac{g^2}{2}\left(\bar{\psi}\psi\right)^2 \Big\}\,, \label{eq:S}
\ee
where $\int_{\tau}\equiv\int_0^{\beta} d\tau$, $\int_x\equiv \int dx$, $\beta=1/T$ is the inverse temperature,
and~$\mu$ is the chemical potential.

We tacitly assume that the field~$\psi$ {in Eq.~\eqref{eq:S}} is built up of $N$ {two-component} spinors,
$\psi^{T}=(\psi_1,\dots,\psi_N)$, with the lower index labeling the $N$ flavors.
Our Euclidean conventions for the Dirac $\gamma$-matrices are 
\be
\gamma_0=\begin{pmatrix} 0 & 1 \\ 1 & 0 \end{pmatrix}\,,\qquad\text{and}\qquad
\gamma_1=\begin{pmatrix} 0 & -{\rm i} \\ {\rm i} & 0 \end{pmatrix}\,,
\ee
with $\{\gamma_{\mu},\gamma_{\nu}\}=2\delta_{\mu\nu}\mathbb{1}_{2\times2}$ and~$\gamma_5={\rm i}\gamma_0\gamma_1$.

{In $d=1+1$}, the GN model is perturbatively renormalizable and asymptotically free.
{The} coupling~$g$ is dimensionless and represents a {marginally relevant} parameter
at the Gau\ss ian fixed point. Since we do not {account} for a {bare} fermion mass in the present study, 
the value of any {dimensionful} physical quantity~$\mathcal O$ only depends on our choice for the value of the
coupling~$g$ at a given UV {momentum} scale~$\Lambda$. This implies that all physical observables are uniquely fixed by
our choice for the coupling. Here, we choose the {physical} fermion mass~$M_0$ at 
vanishing temperature and chemical potential to set the scale. Dimensionless {ratios}~$\mathcal O/M_0$ are then
given by {\it universal} numbers, i.e. they are independent of our actual choice for~$g$. 

{The action~\eqref{eq:S} of the GN model is invariant under global U$(N)$ transformations
of the fermion fields,}
implying that
the associated U($1$) charge is conserved for each flavor separately. {It} is 
also invariant under discrete $\mathbb{Z}_2$ chiral transformations:
\be
\bar{\psi}\mapsto -\bar{\psi}\gamma_5\,,\qquad \psi \mapsto \gamma_5\psi\,{,} \label{eq:trafo}
\ee
{implying that $\bar\psi\psi\mapsto-\bar\psi\psi$.}
However, the infrared regime (long-range limit) of the theory is governed by 
dynamical chiral symmetry breaking, independent of our choice for the coupling~$g$, see also below.

The chiral symmetry of the model can be associated with a $\mathbb{Z}_2$ symmetry {of} the 
{scalar} order parameter~$\langle\bar{\psi}\psi\rangle$.
To see this explicitly, it is convenient to introduce a real-valued auxiliary {scalar} field~$\sigma$ into the path integral 
{representation of the partition function} ${\mathcal Z}$,
\be
{\mathcal Z}=\int {\mathcal D}\bar{\psi}{\mathcal D}\psi\,{\rm e}^{-S}\,, \label{eq:Z}
\ee
by means of a Hubbard-Stratonovich transformation~\cite{Hubbard:1959ub,Stratonovich}. Formally, this is done by multiplying~${\mathcal Z}$ with 
a suitably chosen Gau\ss ian factor,
\be
1={\mathcal N}\int {{\mathcal D}\sigma}\,{\rm e}^{-\frac{1}{2g^2}\int_{\tau}\int_x \sigma^2 }\,,\label{eq:ht1}
\ee
where~$\mathcal N$ is a normalization factor, and then shifting the auxiliary field as follows:
\be 
\sigma \mapsto \sigma + {\rm i}g^2(\bar{\psi}\psi)\,.\label{eq:sigmatrafo}
\ee
The field~$\sigma$ carries the same quantum numbers as the composite field~$\bar{\psi}\psi$. Its equation
of motion is trivially given by $\sigma = - {\rm i}(\bar{\psi}\psi)$, such that {obviously}
$\sigma \mapsto -\sigma$ under discrete chiral $\mathbb{Z}_2$ transformations. 

{Employing Eqs.~\eqref{eq:ht1} and~\eqref{eq:sigmatrafo}, the} action in Eq.~\eqref{eq:Z} changes as $S\mapsto S_{\rm B}$, with the
associated so-called partially bosonized action of the GN model {given by}
\be
  S_{\rm B} 
  =\int_{\tau} \int_x\, \Big\{\bar\psi\left({\rm i}\slashed{\partial} + {\rm i}\sigma + {\rm i}\mu\gamma_0\right)\psi 
  + \frac{1}{2g^2}\sigma^2\Big\}\,.\label{eq:bosact}
\ee
As the fermion {fields} appear only bilinearly in the action, they can be integrated out exactly, leaving us with a 
highly non-local purely bosonic effective action, {see also our discussion below}. 

{Subsequently, we shall split up} the $\sigma$ field into a {background field}~$\bar{\sigma}\equiv\langle\sigma\rangle$
and a fluctuation field, $\sigma\mapsto \bar{\sigma}+\sigma$. The action~$S_{\rm B}$ can then be expanded in powers
of the bosonic fluctuation field. As we are only interested in a study of the large-$N$ limit {here}, it suffices to {restrict} ourselves to the zeroth
order of this expansion, i.e., to simply substitute $\sigma$ with $\bar\sigma$.
Diagrammatically, within the framework of the partially bosonized theory~\eqref{eq:bosact}, this means that Feynman diagrams with internal boson
lines are not taken into account. 
Higher orders in an expansion of the fluctuation field $\sigma$ are suppressed parametrically by powers of~$1/N$ \cite{Pausch:1991ee}. The associated 
Feynman diagrams contain at least one internal boson line.

{It is then {apparent} that the action~\eqref{eq:bosact} with $\sigma\mapsto\bar\sigma$ is invariant under the 
discrete $\mathbb{Z}_2$ chiral transformations~\eqref{eq:trafo} for $\bar\sigma=0$, but not for $\bar\sigma\neq0$,
confirming that $\bar\sigma$ and thus also $\langle\bar\psi\psi\rangle$ constitutes an order parameter for chiral symmetry breaking.}

\subsection{Fermion Doubling}\label{sec:fd}
Let us now discuss our fermion doubling approach. To this end, we discuss the computation of the effective order-parameter potential
of the GN model in the large-$N$ limit and show that our approach allows us to search for the emergence of inhomogeneous phases
in an efficient way. 

For our study, it is convenient to switch to momentum space by using the Fourier representation of the fermion fields:
\be
 \psi(x)&=&\sum_n\int_p e^{-{\rm i}(\nu_n\tau+px)}\psi_n(p)\,,\\  
 \bar\psi(x)&=&\sum_n\int_p e^{{\rm i}(\nu_n\tau+px)}\bar\psi_n(p)\,,
\ee
where~$\int_p\equiv \int \frac{dp}{2\pi}$ and $\nu_n=(2n+1)\pi T$ are the fermionic Matsubara frequencies. 
For the real-valued background field~$\bar{\sigma}$, we employ the following 
ansatz:
\be
 \bar{\sigma}(x)=M\cos(2Q x)=\frac{M}{2}\left(e^{2{\rm i}Qx}+e^{-2{\rm i}Qx}\right)\,.\label{eq:sigans}
\ee
Here, $M\geq 0$ and~$Q$ are real-valued parameters. Chiral symmetry breaking is associated
with a ground-state configuration with {finite $M$}. If the ground-state configuration
assumes a finite value~$Q$, then translation invariance is broken spontaneously. Thus,~$Q$ can be viewed
as an order-parameter for translation symmetry breaking. {\it A priori}, both
types of symmetry breaking are not related.

We add that the ansatz~\eqref{eq:sigans} corresponds to an inhomogeneity of the Larkin-Ovchinnikov-type and includes 
the case of a homogeneous ground state in the limit~$Q\to 0$.
In many-body physics, the quantity~$Q$ can be related
to the center-of-mass momentum of the two-body bound-state associated with condensation. 

Our ansatz for~$\bar{\sigma}$ {shares important properties} with the analytic solution in the large-$N$ limit:
In Refs.~\cite{Thies:2003kk,Schnetz:2004vr}, it was indeed shown that the {true} inhomogeneous ground state 
is described by a periodic function. 
At finite temperature close to the chiral phase boundary, 
it was moreover found that the functional form of the ground state configuration {approaches a single cosine} of the form~\eqref{eq:sigans}.
For small temperatures, on the other hand, our ansatz for~$\bar{\sigma}$ is expected to become insufficient. In fact, the ground-state
solution in the zero-temperature limit is given by a kink-antikink crystal~\cite{Brzoska:2001iq,Thies:2003br,Thies:2003kk,Schnetz:2004vr}, 
also denoted as baryon crystal, which can {obviously} not be described by a simple single-cosine ansatz.

Inserting our ansatz~\eqref{eq:sigans} into the action~\eqref{eq:bosact} yields the following expression:
{\be
S_{\rm B}[\bar{\sigma}] =\Delta S_{\rm B}+ \frac{1}{2g^2}\int_{\tau}\int_x\bar{\sigma}^2\,,
\ee
with}
\be
 \!\!\!\!\!\Delta S_{\rm B}&=&
 \beta\sum_{n}\int_{p}\Big\{\bar\psi_n(p)\left(\nu_n\gamma_0+p\gamma_1+{\rm i}\mu\gamma_0\right)\psi_n(p) \nn\\
  && \!\!\!\!\!\!\!\! +\, \frac{M}{2}\left(\bar\psi_n(p\!+\!2Q)\psi_n(p)+\bar\psi_n(p\!-\!2Q)\psi_n(p)\right)\!\Big\}\,{.} \label{eq:DeltaS}
  \\
  \nn
\ee
Focussing on the momentum structure, we observe that the action can be rewritten as follows:
\begin{widetext}
\be
 \Delta S_{\rm B}&=&\frac{\beta}{2}
 \sum_n\int_p
 \Big\{\bar\psi_n(p-Q)\gamma_0\left(\nu_n + \gamma_0\gamma_1(p-Q)+{\rm i}\mu\right)\psi_n(p-Q)+M\bar\psi_n(p+Q)\gamma_0\gamma_0\psi_n(p-Q) \nn\\
 && \qquad\qquad\qquad +\, \bar\psi_n(p+Q)\gamma_0\left(\nu_n +\gamma_0\gamma_1(p+Q)+{\rm i}\mu\right)\psi_n(p+Q)+M\bar\psi_n(p-Q)\gamma_0\gamma_0\psi_n(p+Q)\Big\}\,.
\ee
\end{widetext}
Here, we have shifted $p$ to $p-Q$ in the first line and $p$ to $p+Q$ in the second line.
For convenience, we now introduce an auxiliary field vector~$\Psi$,
\be
 \Psi_n=\begin{pmatrix}\psi_n(p-Q) \\ \psi_n(p+Q)\end{pmatrix}\,,
\ee
which allows us to rewrite the action {compactly} as follows:
\be
 \Delta S_{\rm B}=\frac{\beta}{2}\sum_n\int_p\Psi_n^\dagger \Delta S^{(2)}_{\rm B}\Psi_n\,.\label{eq:gnfinal}
\ee
The matrix $\Delta S^{(2)}_{\rm B}$ is given by
\be
 \Delta S_{\rm B}^{(2)}=\begin{pmatrix}G_{+}(-Q)& 0 & 0 & {\rm i}M\\0&G_{-}(-Q)&{\rm i}M&0\\&{\rm i}M&G_{+}(Q)&0\\{\rm i}M&0&0&G_{-}(Q)\end{pmatrix}\,,\nn
\ee
where $G_{\pm}({Q})=\nu_n+{\rm i}\mu\pm{\rm i}(p+Q)$. 

Up to this point, we have simply rewritten the action of the GN model.
{However, in the next step} we integrate out the fermion fields by 
assuming that~$\psi(p-Q)$ and~$\psi(p+Q)$ are independent degrees of freedom. 
{Of course, the latter assumption is in general not justified, and will be examined critically below.}
{It} effectively corresponds to a doubling of the number of fermion {fields, which explains why we refer to it as a ``fermion doubling trick"}. 
Setting~$Q=0$, we still {recover} 
the well-known result for the effective mean-field potential {based on the assumption of} a homogeneous ground state~\cite{Wolff:1985av}. 
For~$Q\neq 0$, however,
our doubling prescription is clearly an approximation which we shall discuss 
below in more detail. In any case, for the time being, we {stick to the assumption} that~$\psi(p-Q)$ and~$\psi(p+Q)$ are independent. 

Bearing the doubling issue in mind, we now compute the effective potential~$V$ which is obtained 
from the effective action {defined as}~$\Gamma[\bar{\sigma}]=-\ln {\mathcal Z}$ {and reads}
{\be
\frac{V}{N}&=&\lim_{L\to\infty}\frac{1}{\beta N L}\Gamma[\bar{\sigma}] \nn\\
&=& \frac{1}{2g^2\beta NL }\int_{\tau}\int_x \bar{\sigma}^2
-\frac{1}{2\beta N L}\text{Tr}\ln \Delta S_{\rm B}^{(2)}\,.\label{eq:trlog}
\ee
Here, $L$ denotes the spatial extent of the system.
The trace includes a sum over spin and flavor degrees of freedom. In comparison to the standard calculation, 
a factor of~$2$ appears in the denominator to account for the doubling of the fermionic degrees of freedom.}
With the eigenvalues~$\epsilon_i^{\mp}$ of the matrix~$\Delta S_{\rm B}^{(2)}$,
\be
 &&\epsilon_1^{(\mp)}\!=\! \nu_n\!+\! {\rm i}(\mu\mp E\!-\!Q)
 \quad\text{and}\quad
 \epsilon_2^{(\mp)}\!=\! \nu_n\!+\! {\rm i}(\mu\mp E\!+\!Q)\,,\nn
\ee
where $E=\sqrt{p^2+M^2}$, {the trace} in Eq.~\eqref{eq:trlog} can be computed explicitly. We obtain
\be
 \frac{V}{N}&=&\frac{M^2{\mathcal I}}{2Ng^2} 
   -\Delta V\,, \label{eq:VN}
   \ee
where
\be
{\mathcal I}=\lim_{L\to\infty}\frac{1}{L}\int_{-L/2}^{L/2}dx \cos^2(2Qx)=\begin{cases}
                                                                          1\ \ {\rm for}\ \ Q=0 \\
                                                                          \frac{1}{2}\ \ {\rm for}\ \ Q\neq0
                                                                         \end{cases}\!\!\!\!\!\!, \nonumber
\ee
{and
\be
\Delta V &=&\frac{1}{2\beta}\int_p \Big\{2\beta\left(E+\mu\right)\nn\\
&& {\quad+\sum_{\alpha_1,\alpha_2}\ln\left(1+e^{-\beta(E+\alpha_1Q-\alpha_2\mu)}\right)\Big\}\,} \label{eq:DeltaV}
\ee
with $\alpha_i\in\{-1,1\}$.}
The second term on the right-hand side {of Eq.~\eqref{eq:DeltaV}} is finite and includes the finite-temperature corrections.
On the other hand, the first term is a divergent vacuum contribution.

Next, we renormalize the effective potential in order to cancel all divergent parts.
To this end, {in complete analogy to the calculation for a homogeneous condensate presented in Ref.~\cite{Schon:2000qy},} we consider the potential
in the vacuum limit, i.e.~$T=\mu=0$:
\be
 \frac{V_{0}}{N}&=&\frac{M^2{\mathcal I}}{2Ng^2} -
 \frac{M^2}{4\pi}+\frac{M^2}{2\pi}\ln\left(\frac{M}{\Lambda}\right)\,,\label{eq:vacpot}
\ee
where we have dropped an irrelevant constant independent of~$M$, $Q$, $T$ and $\mu$. 
The vacuum gap equation is obtained
straightforwardly {by minimizing} Eq.~\eqref{eq:vacpot} with respect to~$M$:
\be
\frac{1}{N} \frac{\partial V_{0}}{\partial M}\Big|_{M=M_0}=
 \frac{M_0{\mathcal I}}{Ng^2}\!+\!\frac{M_0}{\pi}\ln\left(\frac{M_0}{\Lambda}\right)\stackrel{!}{=}0\,.\label{eq:gapeq}
\ee
Here, $M_0$ denotes the position of the ground state of the potential~$V_0$ 
and can be identified with the (dynamically)
generated mass of the fermions. 
{Solving the gap equation~\eqref{eq:gapeq} for the coupling $g$, we find:
\be
\frac{\pi{\mathcal I}}{Ng^2}= -\ln\left(\frac{M_0}{\Lambda}\right)\label{eq:M0}\,,
\ee
which, inserted in Eq.~\eqref{eq:VN}, renders the effective potential finite.
For completeness, we note that the so renormalized potential is a continuous function of~$Q$ for~$Q\to 0$, as it should be.}

In the following, we shall consider the so-called `t-Hooft limit, i.e. we keep $Ng^2$ fixed for $N\to \infty$. 
The mass parameter~$M_0$ determines the physical fermion mass and sets the overall scale for all physical
observables. 
We observe that~$g^2$ decreases logarithmically when~$\Lambda$ is increased,~$g^2 \sim 1/\ln(\Lambda)$, as
it should be for an asymptotically free theory.
{As mentioned above,
with} the aid of the relation~\eqref{eq:M0} between the fermion mass~$M_0$ and the coupling~$g$, the effective potential~\eqref{eq:VN}
can be rendered cutoff independent and the limit~$\Lambda\to\infty$ can be safely considered.\footnote{Terms independent of~$M$ 
may still depend on the cutoff. However, this is irrelevant for the present study as such contributions
can be absorbed in a redefinition of the absolute value of the ground-state energy.} Thus, the GN model indeed depends only on a 
single input parameter, e.g. our choice for the fermion mass~$M_0$, and we may therefore choose to measure
all dimensionful physical quantities in units of~$M_0$.

Finally, we would like to give a first critical discussion of our fermion doubling approach. 
{To begin with, we emphasize again that}, by construction, the well-known results for
the {``old"}~$(T,\mu)$ phase diagram {based on the (actually too restrictive) assumption of a homogeneous ground state} 
are recovered~\cite{Wolff:1985av} {(cf. also Sec.~2 of \cite{Thies:2006ti})}, if we set~$Q=0$. 
{Next, we note that the momentum structure of the two-point 
function {{associated with} the $\bar\sigma$ field~\eqref{eq:sigans}} derived from the {fermion} doubling approach agrees with the one {obtained} 
from an exact treatment of the fermion
determinant. This is of utmost importance for
the search for inhomogeneous ground states and will be {explained and} discussed {in detail} in Sect.~\ref{eq:vexp}. 
{The momentum structures of higher $n$-point functions are in general only reproduced approximately.
For the two-point function as obtained from the present approach, however, we still find that its momentum structure 
agrees with the exact one if we take higher Fourier coefficients in our ansatz~\eqref{eq:sigans} into account, see also Eq.~\eqref{eq:genfexp} below.} 
In any case,} a single cosine ansatz 
{for the ground state condensate} seems {to constitute a} natural {choice} for a first search for inhomogeneous 
phases and indeed underlies many analytic or numerical studies of various 
different models, see, e.g., Refs.~\cite{FF,LO,Nickel:2009wj,Kojo:2009ha,Stoof,Radzihovsky,Buballa:2014tba,Roscher:2013cma}. 
For the GN model in 1+1 {dimensions}, 
it has even been shown that the exact solution 
for the ground state condensate approaches the form~\eqref{eq:sigans} close to the 
chiral phase boundary~\cite{Thies:2003kk,Schnetz:2004vr,Schnetz:2005ih}.

\subsection{Vertex Expansion of the Effective Action}\label{eq:vexp}
We now discuss chiral symmetry breaking, with an emphasis on {the} search for 
inhomogeneous phases. In particular, we aim at an analysis of the validity of our fermion doubling
approach in this respect.

In the following we shall assume that the transition from a chirally symmetric phase to a phase with
broken chiral symmetry in the ground state is of second order.\footnote{Note that the arguments presented here hold 
only for a second-order {phase} transition associated with chiral symmetry breaking. For a second-order transition associated with spontaneous translation
symmetry breaking but without a change in the chiral properties of the ground state, our arguments can in general not be applied,
see also Sect.~\ref{sec:pd}.}
For the 1+1 {dimensional} GN model, this has indeed been found
to be the case in the $(T,\mu)$ phase diagram~\cite{Schnetz:2004vr,Schnetz:2005ih}. 

At a second-order chiral {phase} transition, the curvature of the effective potential at~$\bar{\sigma}=0$ changes its sign. 
To be more specific, 
let us consider a vertex expansion of the effective action {about $\bar\sigma=0$}:\footnote{In the {vertex} expansion~{\eqref{eq:GL}}, we 
{do not account for} a dependence of the field~$\sigma$ on the (imaginary) time which would appear in the most general form
of the vertex expansion.}
\be
\Gamma[{\bar{\sigma}}] &=& \sum_{n=1}^{\infty}\frac{1}{(2n)!}\prod_{j=1}^{2n}\int_{x_j}
\Gamma^{(2n)}\bar{\sigma}(x_1)\cdots \bar{\sigma}(x_{2n})\,,\label{eq:GL}
\ee
where we have neglected field-independent terms on the right-hand side.
Here, $\Gamma^{(n)}\equiv \Gamma^{(n)}(x_1,\dots,x_{n})$ is the $n$-th functional derivative of $\Gamma[\bar{\sigma}]$ with respect to the field~$\bar{\sigma}$ evaluated
at~$\bar{\sigma}=0$. For odd~$n$, we have~$\Gamma^{(n)}=0$ due to the chiral symmetry of the GN model.
In order to study the emergence of a chiral condensate as a function of temperature~$T$ and chemical potential~$\mu$, it now
suffices to compute the two-point function~$\Gamma^{(2)}$. With the aid of the GN model, we demonstrate that
our fermion doubling approach indeed {reproduces the correct momentum structure for this function.}

In momentum space, the {leading non-vanishing contribution to the} vertex expansion reads:\footnote{Since we expand~$\Gamma$
about~$\bar{\sigma}=0$, we can make use of the fact that the GN model is a translation-invariant theory. For an
expansion about~$\bar{\sigma}\neq 0$ (e.g. the non-trivial ground state associated with spontaneous 
chiral symmetry breaking), this is in general not the case, see also Sect.~\ref{sec:pd}.}
\be
\Gamma[{\bar{\sigma}}] &=& \frac{1}{2}\int_q \Gamma^{(2)}(q)\bar{\sigma}(-q)\bar{\sigma}(q)+\dots\,.\label{eq:GLmom}
\ee
From an exact treatment of the fermion determinant in
Eq.~\eqref{eq:trlog} (i.e. without making use of our fermion doubling {trick}) with general~$\bar{\sigma}(x)$, 
{we obtain the following} result for~$\Gamma^{(2)}$:
\be
\!\!\!\!\!\! \Gamma^{(2)}(q) &=& \delta\Gamma^{(2)}_{\rm R} \nn\\
&& \; + 2\beta N \int_{p}\frac{1\!-\!n_{\rm F}({p}\!-\!\mu)\!-\!n_{\rm F}(p\!+\!q\!+\!\mu)}{q+2p}\,, \label{eq:2point}
\ee
where~$\delta\Gamma^{(2)}_{\rm R}$ is a suitably chosen counterterm renormalizing
the two-point function,~$\delta\Gamma^{(2)}_{\rm R}=-(N\beta/\pi)\ln(M_0/\Lambda)$,
and~$n_{\rm F}(q)$ denotes the Fermi-Dirac distribution for {free non-interacting} fermions:
\be
n_{\rm F}(q)=\frac{1}{{\rm e}^{\beta q}+1}\,.
\ee
From an expansion of the effective order-parameter potential~\eqref{eq:VN} about~$M=0$ up to order~$M^2$, on the other hand, 
we find
\be
{V}(M,Q)=\frac{1}{2}V^{(2)}(0,Q)M^2+\dots\,,\label{eq:potexp}
\ee
{where we have dropped $M$-independent terms and
\be
V^{(2)}&=&\frac{{\mathcal I}}{g^2} + N\int_p\frac{1}{2p}\Big(2\!-\! n_{\rm F}(p\!+\!Q\!+\!\mu)\! -\! n_{\rm F}(p\!-\!Q\!+\!\mu) \nn\\
&& 
\qquad\qquad - n_{\rm F}(p\!+\!Q\!-\!\mu) \!-\! n_{\rm F}(p\!-\!Q\!-\!\mu)\Big)\,.\label{eq:v2dt}
\ee
{Here, we made use of the fact} that the integrand is invariant under~$p\to -p$.
Inserting the ansatz~\eqref{eq:sigans} into Eq.~\eqref{eq:GLmom}, with the two-point function given by Eq.~\eqref{eq:2point}, 
we find that the coefficient~$V^{(2)}$ in Eq.~\eqref{eq:v2dt}
derived from our fermion doubling trick agrees identically {with the one from the} 
exact expansion~\eqref{eq:GLmom} for~$Q=0$ and up to an overall factor of~$2$ for~$Q\neq 0$. Recall that
\be
V=\lim_{L\to\infty}\frac{1}{\beta L}\Gamma\,.
\ee
This implies that, 
{provided the limit~$L\to\infty$ is considered, {the general} momentum structures of the 
two-point function $\Gamma^{(2)}$ as obtained from the two approaches agree.}
This suffices to detect rigorously a sign change in~$\Gamma^{(2)}$ for any~$Q$. {We therefore} 
conclude that our approach based on {the fermion doubling trick} allows us to detect the onset of spontaneous 
chiral symmetry breaking, assuming that the phase transition is of second order. In particular, this remains true, even if the 
associated chiral condensate is inhomogeneous.}

{Taking into account higher orders in the Fourier-cosine expansion of our ansatz~\eqref{eq:sigans}
for~$\bar{\sigma}$, the effective potential~$V$ can in principle be computed along the
lines of Sect.~\ref{sec:fd} {with} the number of fermions {multiplied} accordingly.
We add that the general momentum structure of the two-point function is then 
still recovered correctly. To be more specific, for~$Q=0$, 
the coefficient~$V^{(2)}$ in Eq.~\eqref{eq:v2dt} agrees again identically with the exact result. For finite~$Q$,
on the other hand, the result from our present approach is only correct up to an overall $Q$-independent 
factor of~$2N_{\sigma}$, 
where~$N_{\sigma}$ denotes the truncation order of the Fourier-cosine expansion of the field~$\bar{\sigma}$. 
{However, note that the momentum structure of higher $n$-point functions is only
recovered approximately.}

A few comments are in order. For illustration purposes, we have 
essentially only shown that our fermion doubling approach {reproduces the correct momentum structure 
of the two-point function}
for the GN model {in $1+1$ dimensions}. At no point in this analysis, 
however, we have {made use of the fact} that we are only considering the case of one spatial dimension. Therefore, our 
arguments {presented above} {should also hold for} 
the GN model in {higher dimensions}, at least as long as we only allow for one-dimensional modulations of the ground-state
configuration and restrict ourselves to a single-cosine ansatz.  With the same line of arguments and also with the same restrictions, 
our fermion doubling approach can be applied to other fermionic theories, provided that their action can be {represented}
in the form~\eqref{eq:gnfinal}. This should also include
Nambu-Jona-Lasinio-type models \cite{Nambu:1961tp,Nambu:1961fr} which are often 
used as effective low-energy models {for QCD}.

Finally, we would like to add a comment on standard derivative expansions. The vertex expansion~\eqref{eq:GL} can be cast into
a Ginzburg-Landau (GL) expansion of the following form:
\be
\!\!\!\!\!\!\!\!\!\!\Gamma_{\text{GL}}[\bar{\sigma}]&=&\int_x \Big\{\frac{1}{2!}\Gamma_{2,0}\bar{\sigma}^2 + \frac{1}{4!}\Gamma_{4,0} \bar{\sigma}^4 +\dots \nn\\
&& \quad\; + \frac{1}{2!}\Gamma_{2,2}(\partial_x \bar{\sigma})^2 + \frac{1}{4!}\Gamma_{2,4} (\partial_x^2 \bar{\sigma})^2 +\dots
\Big\}\,,\label{eq:GL2}
\ee
where we have again dropped field-independent terms and already made use of the symmetries of the GN model. The quantities $\Gamma_{i,j}$ denote 
{\it a priori} unknown expansion coefficients which depend {on temperature and chemical potential.}
The indices~$i$ and~$j$ are associated with powers of the field and derivatives thereof,
respectively. For example, the coefficients~$\Gamma_{2,j}$ with $j\geq 0$ effectively span the above studied two-point function.
If we assume~$\bar{\sigma}$ to be homogeneous from 
the beginning, then we have~$\Gamma_{i,j}=0$ for $j>0$. 

For the GN model, we may employ a (general) Fourier cosine series as an ansatz for~$\bar{\sigma}$,
\be
\bar{\sigma}(x)=\sum_{n=0}^{\infty} M^{(n)} \cos(2Qnx)\,.\label{eq:genfexp}
\ee
The derivatives in Eq.~\eqref{eq:GL2} are then effectively translated
into powers of the momentum~$Q$.
The ground state is {then} obtained by a minimization of~$\Gamma_{\text{GL}}$ with respect to the 
parameter~$Q$ and {the Fourier {coefficients~$M^{(n)}$} {with $n\in\mathbb{N}_0$}.}
In the present work, we have {exclusively limited ourselves to} 
a {single-cosine} ansatz for~$\bar{\sigma}$, see Eq.~\eqref{eq:sigans}. However, already with this ansatz, we take into account arbitrarily high 
orders in~$Q$ in the derivative expansion in Eq.~\eqref{eq:GL2}. 
With respect to a reliable prediction of the chiral phase boundary,
any finite order truncation of this power series in~$Q$ is bound to fail. 
Indeed, the present analysis of the two-point function suggests that a low-order expansion in~$Q$
is particularly inadequate to compute the phase boundary if the chiral phase transition {for a given value of~$\mu$} occurs at low 
temperatures.\footnote{Note that such an expansion {is expected to yield an asymptotic series in~$Q$}.}
This observation is in line with Refs.~\cite{Basar:2008ki,Basar:2009fg} and, in our case, it can be traced back to the fact that
the momentum dependence of the two-point function is essentially determined by Fermi-Dirac distributions which turn into
{Heaviside} step functions in the zero-temperature limit. 

\section{Phase Diagram}\label{sec:pd}
The phase diagram of the 1+1 {dimensional} GN model has been discussed in great detail in Refs.~\cite{Schnetz:2004vr,Schnetz:2005ih}
where also the exact solutions for the phase boundaries in the large-$N$ limit can be found. Here, we use these results
to test our fermion doubling approach on a quantitative level.
\begin{figure}[t]
\includegraphics[width=0.97\columnwidth]{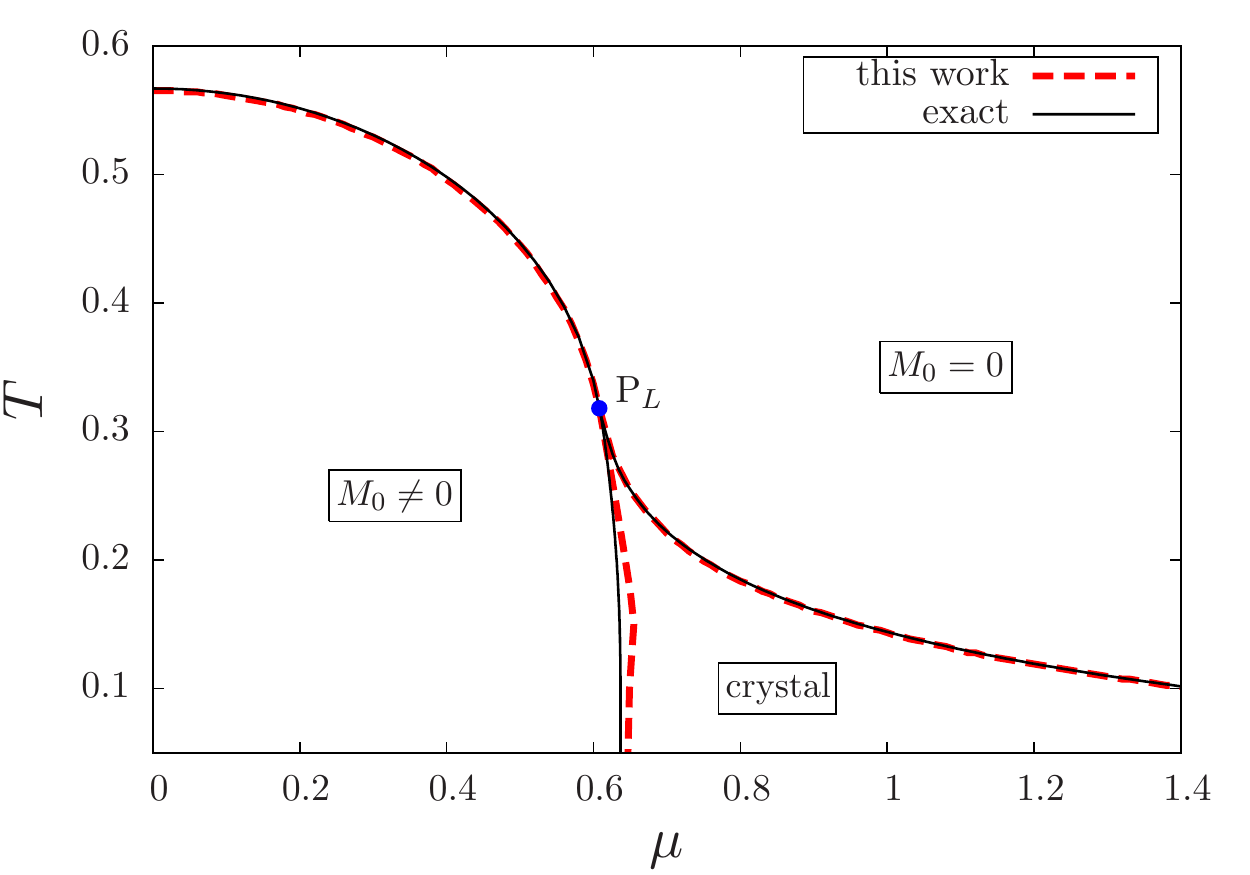}
\caption{\label{Fig:1}(color online) Phase diagram of the GN model {in $1+1$ dimensions} in the $(T,\mu)$ plane. The temperature~$T$ and 
the chemical potential~$\mu$ are measured in units of {the {vacuum} fermion mass}. Our results 
for the phase boundary between the chirally symmetric phase ($M_0=0$) and the phases with spontaneously broken chiral symmetry
agree with the exact solution~{\cite{Thies:2003kk,Schnetz:2004vr,Schnetz:2005ih}}. 
{The boundary between the phase with a homogeneous chiral condensate ($M_0\neq 0,Q=0$) and the crystal phase~($M_0\neq 0,Q\neq 0$)
is not reproduced correctly within our present study,}
see main text for details.}
\end{figure}

In Fig.~\ref{Fig:1}, we show the phase diagram in the $(T,\mu)$ plane. The solid black lines depict the exact solutions for the 
phase boundaries~\cite{Schnetz:2004vr,Schnetz:2005ih}. Overall, three different phases are found to exist: a chirally
symmetric phase ($M_0=0$), a phase with spontaneously broken chiral symmetry described by a homogeneous ground
state ($M_0\neq 0,Q=0$), and an inhomogeneous/crystal phase $(M_0\neq 0,Q\neq 0)$. The latter is {characterized} by 
a spontaneous breakdown of the chiral symmetry as well as of the translation symmetry. Strictly speaking,
there is still a residual discrete translation symmetry as the ground-state is described by a periodic function 
for all values of~$T$ and~$\mu$.

It was found in Refs.~\cite{Thies:2003kk,Schnetz:2004vr}
that the three phases are separated by second-order 
phase transitions. The associated phase transition lines meet at a {\it Lifshitz} point. Note that both the phase with a homogeneous condensate
and the one with an inhomogeneous condensate are characterized by chiral symmetry breaking. The boundary between these two phases
is solely associated with translation symmetry breaking where the value of $Q$ of the ground-state configuration acts as an order parameter.
{We add here that within the phase with a {nonvanishing} homogeneous condensate, 
{there exists} a so-called metastable phase in which the order-parameter potential, 
apart from an absolute minimum at $\bar{\sigma}\neq0$,} acquires a 
{local} minimum at~$\bar{\sigma}=0$, see Ref.~\cite{Wolff:1985av}. 
{Note also that the} originally found first-order transition line 
{-- obtained by allowing for homogeneous condensates only -- between the phases with finite and vanishing homogeneous condensates~\cite{Wolff:1985av}} lies
within the crystal phase. In the {correct} phase diagram~{\cite{Thies:2003kk,Schnetz:2004vr}} {depicted in Fig.~\ref{Fig:1}}, however, this line does not describe
a phase transition in the ground-state of the theory anymore.

Let us now discuss the results from a minimization of the effective order-parameter potential~$V$, see Eq.~\eqref{eq:VN}.
In perfect agreement with our analytic analysis of the fermion doubling approach in Sects.~\ref{sec:fd} and~\ref{eq:vexp}, we find that
the {boundary} between the phases with broken chiral symmetry and the
chirally symmetric phase {is} recovered correctly. The position of the {\it Lifshitz} point, denoted by $P_L$ in Fig.~\ref{Fig:1}, 
{is reproduced correctly} as well, {within our numerical errors}.
Moreover, we would like to {emphasize} that the dependence of~$Q$ on~$\mu$ along the transition line between the crystal phase and the
chirally symmetric phase is in perfect agreement with the exact result for all studied values of~$\mu$. In particular, we find
that $Q(\mu)$ tends to zero continuously when the {\it Lifshitz} point is approached. For large~$\mu\gg M_0$, on the other hand, we have~$Q(\mu)\sim \mu$,
as naively expected from a dimensional analysis.\footnote{It is worth emphasizing that the transition between the crystal phase
and the chirally symmetric phase describes a first-order transition line for translation-symmetry breaking but a second-order line
for chiral symmetry breaking.}

Our result for the {boundary} between the phase with a homogeneous condensate and the crystal phase differs
from the exact solution~\cite{Thies:2003kk,Schnetz:2005ih}. To be more specific, close to the {\it Lifshitz} point, we still
find that our fermion doubling approach yields the correct result for the transition line. Decreasing the temperature, however,
we observe that our result starts to deviate from the exact solution. 
{Remarkably, as can be seen from Fig.~\ref{Fig:1}, it is still well-compatible with the exact phase boundary.}
{In any case, also with our present approach, we obtain that the originally found first-order transition line lies still
within the crystal phase.}

At this point, we recall that the transition between the phase with a homogeneous condensate and the crystal phase is not related to
chiral symmetry breaking. {The respective phase boundary} only signals translation symmetry breaking as measured by the value of~$Q$ of the ground-state configuration.
The associated phase transition was found to be of second order in the exact solution of the GN model~\cite{Thies:2003kk,Schnetz:2004vr}. 
With our fermion doubling approach, we instead find a weak first-order transition. 
The discrepancy between the exact solution and our results in the position and the nature of the transition line can be traced back to the fact that
the ground-state of the crystal phase approaches a so-called kink-antikink solution in the 
zero-temperature limit~\cite{Brzoska:2001iq,Thies:2003br,Thies:2003kk}. Thus, higher orders in the Fourier decomposition of the ground state 
configuration are expected to become {increasingly} important when the temperature is decreased and our simple single cosine 
ansatz for~$\bar{\sigma}$ does not represent an adequate approximation anymore.
Moreover, we emphasize again that our arguments {concerning the possibility of a reliable determination of the phase boundaries} in Sect.~\ref{eq:vexp} hold only for 
second-order chiral
phase transitions for which the curvature of the effective potential at~$\bar{\sigma}=0$ serves {as} an order parameter. The latter is not the 
case {for the} transition between a crystal phase and a homogeneous phase, where both {phases} are governed by chiral symmetry breaking. 
Here, higher $n$-point functions may indeed become relevant. In the present case, in particular {for the GN model in $1+1$ dimensions}, the importance of higher $n$-point functions 
is {to be} expected since the order-parameter potential acquires an additional {{local -- but not global} --} 
minimum at~$\bar{\sigma}=0$ in the phase with a homogeneous condensate close
to the transition to the crystal phase, see also our discussion above.

{In contrast to the momentum structure of
the first non-trivial expansion coefficient~$V^{(2)}$ of the order-parameter potential (see Eq.~\eqref{eq:potexp}),
the ones for the higher-order coefficients {determined} from our {fermion doubling} approach do {in general} not agree 
with those from an exact treatment of the fermion determinant. Still, the results for all expansion coefficients
from our present approach are identical to the exact ones for~$Q=0$.\footnote{Recall that 
the expansion coefficients of the potential are intimately connected with the
$n$-point functions, see also our discussion in Sect.~\ref{eq:vexp}.} 
Since} the transition line to the crystal phase 
is of second order in~$Q$~\cite{Schnetz:2004vr,Schnetz:2005ih} implying that the homogeneous
ground state is continuously connected to the inhomogeneous one, it is still reasonable to 
expect that our fermion doubling approach allows to give a reliable first estimate for the position of the transition line, provided that the exact ground-state
can be well described by a simple ansatz of the form~\eqref{eq:sigans}. For the 1+1 {dimensional} GN model, this is indeed the case in the vicinity of the chiral phase 
boundary~\cite{Schnetz:2004vr,Schnetz:2005ih} {and explains} why our result for the transition line {from the phase with $M_0\neq0$ and $Q=0$} to the crystal phase 
is still in {excellent} agreement with the exact solution {close to the {\it Lifshitz} point}. {As discussed above, for} lower temperatures,
our simple ansatz for~$\bar{\sigma}$ does no longer provide an adequate description of the ground state and the agreement with the exact solution
for the transition line is only qualitative, see Fig.~\ref{Fig:1}. 
In any case, the overall agreement of our results {based on the fermion doubling trick} with the exact results is quite
impressive {for the entire phase diagram}, {in particular} given the simplicity of our fermion doubling approach.

\section{Conclusions}\label{sec:conc}
In the present work we have introduced and critically discussed a simple ``fermion doubling trick" which allows us
to search for the emergence of inhomogeneous phases in the phase diagram of fermionic models. Our approach is efficient
in the sense that {it} is based on a straightforward minimization of the effective order-parameter potential ({of complexity} analogous to
standard mean-field studies {assuming homogeneous condensates}) and does not require any exact diagonalization methods, such as {the solution of}
{\it Bogoliubov-de Gennes}-type equations. Exemplarily employing {our fermion doubling trick for} the $d=$1+1 {dimensional} GN model, we have indeed found that 
the chiral phase boundary agrees identically with the exact solution in Refs.~\cite{Thies:2003kk,Schnetz:2004vr,Schnetz:2005ih}, including
the prediction of the emergence of a crystal phase. Moreover,
our result for the transition line between the phase with a homogeneous chiral condensate and the crystal phase approaches
the exact solution close to the {\it Lifshitz} point and agrees at least qualitatively with the exact solution for low temperatures.

In order to show that the chiral phase boundary is indeed predicted correctly by our fermion doubling approach, we have analyzed
the two-point function and found that {its momentum structure agrees} 
with the one from an exact computation. In addition, we have argued that this is not only the case
for the GN model {in $1+1$ dimensions} but also {expected}
for {Nambu-Jona-Lasinio}-type models and also applies to these types of models in higher dimensions, as long as only
one-dimensional modulations of the ground state are considered. Apart from the chiral phase boundary, we have also pointed out
why our fermion doubling approach may still yield a reasonable first estimate for the transition line between the 
phase with a homogeneous chiral condensate and the crystal phase for a given model.
From our discussion, it is also clear that our present approach is not capable of determining the exact ground-state energy,
in particular within the crystal phase. By construction, the strength of {the} approach is rather to detect the emergence of
crystal phases in the phase diagram of fermionic theories in a comparatively simple and efficient way 
and thereby to draw a more detailed picture of the {interaction} dynamics
underlying these theories. In this respect, our fermion doubling approach may help to guide future phase diagram
studies and assist as well as direct other more powerful (exact) methods, such as {exact diagonalization},
in the search for the ground state of fermionic models.

{\it Acknowledgments.--~} The authors thank G.~V.~Dunne, S.~Rechenberger and M.~Thies for useful discussions and 
M.~Thies also for useful comments on the manuscript.
J.B. and D.R. acknowledge support
by HIC for FAIR within the LOEWE program of the State of Hesse as well as by the DFG under Grant BR 4005/2-1.


%
\bibliography{gn}

\begin{thebibliography}{45}%
\makeatletter
\providecommand \@ifxundefined [1]{%
 \@ifx{#1\undefined}
}%
\providecommand \@ifnum [1]{%
 \ifnum #1\expandafter \@firstoftwo
 \else \expandafter \@secondoftwo
 \fi
}%
\providecommand \@ifx [1]{%
 \ifx #1\expandafter \@firstoftwo
 \else \expandafter \@secondoftwo
 \fi
}%
\providecommand \natexlab [1]{#1}%
\providecommand \enquote  [1]{``#1''}%
\providecommand \bibnamefont  [1]{#1}%
\providecommand \bibfnamefont [1]{#1}%
\providecommand \citenamefont [1]{#1}%
\providecommand \href@noop [0]{\@secondoftwo}%
\providecommand \href [0]{\begingroup \@sanitize@url \@href}%
\providecommand \@href[1]{\@@startlink{#1}\@@href}%
\providecommand \@@href[1]{\endgroup#1\@@endlink}%
\providecommand \@sanitize@url [0]{\catcode `\\12\catcode `\$12\catcode
  `\&12\catcode `\#12\catcode `\^12\catcode `\_12\catcode `\%12\relax}%
\providecommand \@@startlink[1]{}%
\providecommand \@@endlink[0]{}%
\providecommand \url  [0]{\begingroup\@sanitize@url \@url }%
\providecommand \@url [1]{\endgroup\@href {#1}{\urlprefix }}%
\providecommand \urlprefix  [0]{URL }%
\providecommand \Eprint [0]{\href }%
\providecommand \doibase [0]{http://dx.doi.org/}%
\providecommand \selectlanguage [0]{\@gobble}%
\providecommand \bibinfo  [0]{\@secondoftwo}%
\providecommand \bibfield  [0]{\@secondoftwo}%
\providecommand \translation [1]{[#1]}%
\providecommand \BibitemOpen [0]{}%
\providecommand \bibitemStop [0]{}%
\providecommand \bibitemNoStop [0]{.\EOS\space}%
\providecommand \EOS [0]{\spacefactor3000\relax}%
\providecommand \BibitemShut  [1]{\csname bibitem#1\endcsname}%
\let\auto@bib@innerbib\@empty
\bibitem [{\citenamefont {Fulde}\ and\ \citenamefont {Ferrell}(1964)}]{FF}%
  \BibitemOpen
  \bibfield  {author} {\bibinfo {author} {\bibfnamefont {P.}~\bibnamefont
  {Fulde}}\ and\ \bibinfo {author} {\bibfnamefont {R.~A.}\ \bibnamefont
  {Ferrell}},\ }\href@noop {} {\bibfield  {journal} {\bibinfo  {journal} {Phys.
  Rev.}\ }\textbf {\bibinfo {volume} {135}},\ \bibinfo {pages} {A550} (\bibinfo
  {year} {1964})}\BibitemShut {NoStop}%
\bibitem [{\citenamefont {Larkin}\ and\ \citenamefont
  {Ovchinnikov}(1965)}]{LO}%
  \BibitemOpen
  \bibfield  {author} {\bibinfo {author} {\bibfnamefont {A.~I.}\ \bibnamefont
  {Larkin}}\ and\ \bibinfo {author} {\bibfnamefont {Y.~N.}\ \bibnamefont
  {Ovchinnikov}},\ }\href@noop {} {\bibfield  {journal} {\bibinfo  {journal}
  {Sov. Phys. JETP}\ }\textbf {\bibinfo {volume} {20}},\ \bibinfo {pages} {762}
  (\bibinfo {year} {1965})}\BibitemShut {NoStop}%
\bibitem [{\citenamefont {Jackiw}\ and\ \citenamefont
  {Schrieffer}(1981)}]{Jackiw:1981wc}%
  \BibitemOpen
  \bibfield  {author} {\bibinfo {author} {\bibfnamefont {R.}~\bibnamefont
  {Jackiw}}\ and\ \bibinfo {author} {\bibfnamefont {J.}~\bibnamefont
  {Schrieffer}},\ }\href {\doibase 10.1016/0550-3213(81)90557-5} {\bibfield
  {journal} {\bibinfo  {journal} {Nucl.Phys.}\ }\textbf {\bibinfo {volume}
  {B190}},\ \bibinfo {pages} {253} (\bibinfo {year} {1981})}\BibitemShut
  {NoStop}%
\bibitem [{\citenamefont {Mertsching}\ and\ \citenamefont
  {Fischbeck}(1981)}]{Mertsching}%
  \BibitemOpen
  \bibfield  {author} {\bibinfo {author} {\bibfnamefont {J.}~\bibnamefont
  {Mertsching}}\ and\ \bibinfo {author} {\bibfnamefont {H.~J.}\ \bibnamefont
  {Fischbeck}},\ }\href@noop {} {\bibfield  {journal} {\bibinfo  {journal}
  {phys. stat. sol. (b)}\ }\textbf {\bibinfo {volume} {103}},\ \bibinfo {pages}
  {783} (\bibinfo {year} {1981})}\BibitemShut {NoStop}%
\bibitem [{\citenamefont {Machida}\ and\ \citenamefont
  {Nakanishi}(1984)}]{Machida}%
  \BibitemOpen
  \bibfield  {author} {\bibinfo {author} {\bibfnamefont {K.}~\bibnamefont
  {Machida}}\ and\ \bibinfo {author} {\bibfnamefont {H.}~\bibnamefont
  {Nakanishi}},\ }\href@noop {} {\bibfield  {journal} {\bibinfo  {journal}
  {Phys. Rev.}\ }\textbf {\bibinfo {volume} {B 30}},\ \bibinfo {pages} {122}
  (\bibinfo {year} {1984})}\BibitemShut {NoStop}%
\bibitem [{\citenamefont {Chodos}\ \emph {et~al.}(1999)\citenamefont {Chodos},
  \citenamefont {Minakata},\ and\ \citenamefont {Cooper}}]{Chodos:1998wg}%
  \BibitemOpen
  \bibfield  {author} {\bibinfo {author} {\bibfnamefont {A.}~\bibnamefont
  {Chodos}}, \bibinfo {author} {\bibfnamefont {H.}~\bibnamefont {Minakata}}, \
  and\ \bibinfo {author} {\bibfnamefont {F.}~\bibnamefont {Cooper}},\ }\href
  {\doibase 10.1016/S0370-2693(99)00084-2} {\bibfield  {journal} {\bibinfo
  {journal} {Phys.Lett.}\ }\textbf {\bibinfo {volume} {B449}},\ \bibinfo
  {pages} {260} (\bibinfo {year} {1999})},\ \Eprint
  {http://arxiv.org/abs/hep-ph/9812305} {arXiv:hep-ph/9812305 [hep-ph]}
  \BibitemShut {NoStop}%
\bibitem [{\citenamefont {Kleinert}\ and\ \citenamefont
  {Babaev}(1998)}]{Kleinert:1998kj}%
  \BibitemOpen
  \bibfield  {author} {\bibinfo {author} {\bibfnamefont {H.}~\bibnamefont
  {Kleinert}}\ and\ \bibinfo {author} {\bibfnamefont {E.}~\bibnamefont
  {Babaev}},\ }\href {\doibase 10.1016/S0370-2693(98)00983-6} {\bibfield
  {journal} {\bibinfo  {journal} {Phys.Lett.}\ }\textbf {\bibinfo {volume}
  {B438}},\ \bibinfo {pages} {311} (\bibinfo {year} {1998})},\ \Eprint
  {http://arxiv.org/abs/hep-th/9809112} {arXiv:hep-th/9809112 [hep-th]}
  \BibitemShut {NoStop}%
\bibitem [{\citenamefont {Bulgac}\ and\ \citenamefont
  {Forbes}(2008)}]{Bulgac:2008tm}%
  \BibitemOpen
  \bibfield  {author} {\bibinfo {author} {\bibfnamefont {A.}~\bibnamefont
  {Bulgac}}\ and\ \bibinfo {author} {\bibfnamefont {M.~M.}\ \bibnamefont
  {Forbes}},\ }\href {\doibase 10.1103/PhysRevLett.101.215301} {\bibfield
  {journal} {\bibinfo  {journal} {Phys.Rev.Lett.}\ }\textbf {\bibinfo {volume}
  {101}},\ \bibinfo {pages} {215301} (\bibinfo {year} {2008})},\ \Eprint
  {http://arxiv.org/abs/0804.3364} {arXiv:0804.3364 [cond-mat.supr-con]}
  \BibitemShut {NoStop}%
\bibitem [{\citenamefont {{Baarsma}}\ and\ \citenamefont {{Stoof}}()}]{Stoof}%
  \BibitemOpen
  \bibfield  {author} {\bibinfo {author} {\bibfnamefont {J.~E.}\ \bibnamefont
  {{Baarsma}}}\ and\ \bibinfo {author} {\bibfnamefont {H.~T.~C.}\ \bibnamefont
  {{Stoof}}},\ }\href@noop {} {\ }\Eprint {http://arxiv.org/abs/1212.5450}
  {arXiv:1212.5450 [cond-mat.quant-gas]} \BibitemShut {NoStop}%
\bibitem [{\citenamefont {Radzihovsky}(2012)}]{Radzihovsky}%
  \BibitemOpen
  \bibfield  {author} {\bibinfo {author} {\bibfnamefont {L.}~\bibnamefont
  {Radzihovsky}},\ }\href@noop {} {\bibfield  {journal} {\bibinfo  {journal}
  {Physica C: Superconductivity}\ }\textbf {\bibinfo {volume} {481}},\ \bibinfo
  {pages} {189} (\bibinfo {year} {2012})}\BibitemShut {NoStop}%
\bibitem [{\citenamefont {Roscher}\ \emph {et~al.}(2014)\citenamefont
  {Roscher}, \citenamefont {Braun},\ and\ \citenamefont
  {Drut}}]{Roscher:2013cma}%
  \BibitemOpen
  \bibfield  {author} {\bibinfo {author} {\bibfnamefont {D.}~\bibnamefont
  {Roscher}}, \bibinfo {author} {\bibfnamefont {J.}~\bibnamefont {Braun}}, \
  and\ \bibinfo {author} {\bibfnamefont {J.~E.}\ \bibnamefont {Drut}},\
  }\href@noop {} {\bibfield  {journal} {\bibinfo  {journal} {Phys. Rev.}\
  }\textbf {\bibinfo {volume} {A89}},\ \bibinfo {pages} {063609} (\bibinfo
  {year} {2014})},\ \Eprint {http://arxiv.org/abs/1311.0179} {arXiv:1311.0179
  [cond-mat.quant-gas]} \BibitemShut {NoStop}%
\bibitem [{\citenamefont {Schon}\ and\ \citenamefont
  {Thies}(2000)}]{Schon:2000he}%
  \BibitemOpen
  \bibfield  {author} {\bibinfo {author} {\bibfnamefont {V.}~\bibnamefont
  {Schon}}\ and\ \bibinfo {author} {\bibfnamefont {M.}~\bibnamefont {Thies}},\
  }\href {\doibase 10.1103/PhysRevD.62.096002} {\bibfield  {journal} {\bibinfo
  {journal} {Phys.Rev.}\ }\textbf {\bibinfo {volume} {D62}},\ \bibinfo {pages}
  {096002} (\bibinfo {year} {2000})},\ \Eprint
  {http://arxiv.org/abs/hep-th/0003195} {arXiv:hep-th/0003195 [hep-th]}
  \BibitemShut {NoStop}%
\bibitem [{\citenamefont {Schon}\ and\ \citenamefont {Thies}()}]{Schon:2000qy}%
  \BibitemOpen
  \bibfield  {author} {\bibinfo {author} {\bibfnamefont {V.}~\bibnamefont
  {Schon}}\ and\ \bibinfo {author} {\bibfnamefont {M.}~\bibnamefont {Thies}},\
  }\href@noop {} {\ }\Eprint {http://arxiv.org/abs/hep-th/0008175}
  {hep-th/0008175 [hep-th]} \BibitemShut {NoStop}%
\bibitem [{\citenamefont {Schnetz}\ \emph {et~al.}(2004)\citenamefont
  {Schnetz}, \citenamefont {Thies},\ and\ \citenamefont
  {Urlichs}}]{Schnetz:2004vr}%
  \BibitemOpen
  \bibfield  {author} {\bibinfo {author} {\bibfnamefont {O.}~\bibnamefont
  {Schnetz}}, \bibinfo {author} {\bibfnamefont {M.}~\bibnamefont {Thies}}, \
  and\ \bibinfo {author} {\bibfnamefont {K.}~\bibnamefont {Urlichs}},\ }\href
  {\doibase 10.1016/j.aop.2004.06.009} {\bibfield  {journal} {\bibinfo
  {journal} {Annals Phys.}\ }\textbf {\bibinfo {volume} {314}},\ \bibinfo
  {pages} {425} (\bibinfo {year} {2004})},\ \Eprint
  {http://arxiv.org/abs/hep-th/0402014} {arXiv:hep-th/0402014 [hep-th]}
  \BibitemShut {NoStop}%
\bibitem [{\citenamefont {Schnetz}\ \emph {et~al.}(2006)\citenamefont
  {Schnetz}, \citenamefont {Thies},\ and\ \citenamefont
  {Urlichs}}]{Schnetz:2005ih}%
  \BibitemOpen
  \bibfield  {author} {\bibinfo {author} {\bibfnamefont {O.}~\bibnamefont
  {Schnetz}}, \bibinfo {author} {\bibfnamefont {M.}~\bibnamefont {Thies}}, \
  and\ \bibinfo {author} {\bibfnamefont {K.}~\bibnamefont {Urlichs}},\ }\href
  {\doibase 10.1016/j.aop.2005.12.007} {\bibfield  {journal} {\bibinfo
  {journal} {Annals Phys.}\ }\textbf {\bibinfo {volume} {321}},\ \bibinfo
  {pages} {2604} (\bibinfo {year} {2006})},\ \Eprint
  {http://arxiv.org/abs/hep-th/0511206} {arXiv:hep-th/0511206 [hep-th]}
  \BibitemShut {NoStop}%
\bibitem [{\citenamefont {Basar}\ and\ \citenamefont
  {Dunne}(2008{\natexlab{a}})}]{Basar:2008im}%
  \BibitemOpen
  \bibfield  {author} {\bibinfo {author} {\bibfnamefont {G.}~\bibnamefont
  {Basar}}\ and\ \bibinfo {author} {\bibfnamefont {G.~V.}\ \bibnamefont
  {Dunne}},\ }\href {\doibase 10.1103/PhysRevLett.100.200404} {\bibfield
  {journal} {\bibinfo  {journal} {Phys. Rev. Lett.}\ }\textbf {\bibinfo
  {volume} {100}},\ \bibinfo {pages} {200404} (\bibinfo {year}
  {2008}{\natexlab{a}})},\ \Eprint {http://arxiv.org/abs/0803.1501}
  {arXiv:0803.1501 [hep-th]} \BibitemShut {NoStop}%
\bibitem [{\citenamefont {Basar}\ and\ \citenamefont
  {Dunne}(2008{\natexlab{b}})}]{Basar:2008ki}%
  \BibitemOpen
  \bibfield  {author} {\bibinfo {author} {\bibfnamefont {G.}~\bibnamefont
  {Basar}}\ and\ \bibinfo {author} {\bibfnamefont {G.~V.}\ \bibnamefont
  {Dunne}},\ }\href {\doibase 10.1103/PhysRevD.78.065022} {\bibfield  {journal}
  {\bibinfo  {journal} {Phys.Rev.}\ }\textbf {\bibinfo {volume} {D78}},\
  \bibinfo {pages} {065022} (\bibinfo {year} {2008}{\natexlab{b}})},\ \Eprint
  {http://arxiv.org/abs/0806.2659} {arXiv:0806.2659 [hep-th]} \BibitemShut
  {NoStop}%
\bibitem [{\citenamefont {Basar}\ \emph {et~al.}(2009)\citenamefont {Basar},
  \citenamefont {Dunne},\ and\ \citenamefont {Thies}}]{Basar:2009fg}%
  \BibitemOpen
  \bibfield  {author} {\bibinfo {author} {\bibfnamefont {G.}~\bibnamefont
  {Basar}}, \bibinfo {author} {\bibfnamefont {G.~V.}\ \bibnamefont {Dunne}}, \
  and\ \bibinfo {author} {\bibfnamefont {M.}~\bibnamefont {Thies}},\ }\href
  {\doibase 10.1103/PhysRevD.79.105012} {\bibfield  {journal} {\bibinfo
  {journal} {Phys.Rev.}\ }\textbf {\bibinfo {volume} {D79}},\ \bibinfo {pages}
  {105012} (\bibinfo {year} {2009})},\ \Eprint {http://arxiv.org/abs/0903.1868}
  {arXiv:0903.1868 [hep-th]} \BibitemShut {NoStop}%
\bibitem [{\citenamefont {Bringoltz}(2007)}]{Bringoltz:2006pz}%
  \BibitemOpen
  \bibfield  {author} {\bibinfo {author} {\bibfnamefont {B.}~\bibnamefont
  {Bringoltz}},\ }\href {\doibase 10.1088/1126-6708/2007/03/016} {\bibfield
  {journal} {\bibinfo  {journal} {JHEP}\ }\textbf {\bibinfo {volume} {0703}},\
  \bibinfo {pages} {016} (\bibinfo {year} {2007})},\ \Eprint
  {http://arxiv.org/abs/hep-lat/0612010} {arXiv:hep-lat/0612010 [hep-lat]}
  \BibitemShut {NoStop}%
\bibitem [{\citenamefont {Bringoltz}(2009)}]{Bringoltz:2009ym}%
  \BibitemOpen
  \bibfield  {author} {\bibinfo {author} {\bibfnamefont {B.}~\bibnamefont
  {Bringoltz}},\ }\href {\doibase 10.1103/PhysRevD.79.125006} {\bibfield
  {journal} {\bibinfo  {journal} {Phys.Rev.}\ }\textbf {\bibinfo {volume}
  {D79}},\ \bibinfo {pages} {125006} (\bibinfo {year} {2009})},\ \Eprint
  {http://arxiv.org/abs/0901.4035} {arXiv:0901.4035 [hep-lat]} \BibitemShut
  {NoStop}%
\bibitem [{\citenamefont {Nickel}(2009)}]{Nickel:2009wj}%
  \BibitemOpen
  \bibfield  {author} {\bibinfo {author} {\bibfnamefont {D.}~\bibnamefont
  {Nickel}},\ }\href {\doibase 10.1103/PhysRevD.80.074025} {\bibfield
  {journal} {\bibinfo  {journal} {Phys. Rev.}\ }\textbf {\bibinfo {volume}
  {D80}},\ \bibinfo {pages} {074025} (\bibinfo {year} {2009})},\ \Eprint
  {http://arxiv.org/abs/0906.5295} {arXiv:0906.5295 [hep-ph]} \BibitemShut
  {NoStop}%
\bibitem [{\citenamefont {Kojo}\ \emph {et~al.}(2010)\citenamefont {Kojo},
  \citenamefont {Hidaka}, \citenamefont {McLerran},\ and\ \citenamefont
  {Pisarski}}]{Kojo:2009ha}%
  \BibitemOpen
  \bibfield  {author} {\bibinfo {author} {\bibfnamefont {T.}~\bibnamefont
  {Kojo}}, \bibinfo {author} {\bibfnamefont {Y.}~\bibnamefont {Hidaka}},
  \bibinfo {author} {\bibfnamefont {L.}~\bibnamefont {McLerran}}, \ and\
  \bibinfo {author} {\bibfnamefont {R.~D.}\ \bibnamefont {Pisarski}},\ }\href
  {\doibase 10.1016/j.nuclphysa.2010.05.053} {\bibfield  {journal} {\bibinfo
  {journal} {Nucl.Phys.}\ }\textbf {\bibinfo {volume} {A843}},\ \bibinfo
  {pages} {37} (\bibinfo {year} {2010})},\ \Eprint
  {http://arxiv.org/abs/0912.3800} {arXiv:0912.3800 [hep-ph]} \BibitemShut
  {NoStop}%
\bibitem [{\citenamefont {Buballa}\ and\ \citenamefont
  {Carignano}()}]{Buballa:2014tba}%
  \BibitemOpen
  \bibfield  {author} {\bibinfo {author} {\bibfnamefont {M.}~\bibnamefont
  {Buballa}}\ and\ \bibinfo {author} {\bibfnamefont {S.}~\bibnamefont
  {Carignano}},\ }\href@noop {} {\ }\Eprint {http://arxiv.org/abs/1406.1367}
  {arXiv:1406.1367 [hep-ph]} \BibitemShut {NoStop}%
\bibitem [{\citenamefont {Casalbuoni}\ and\ \citenamefont
  {Nardulli}(2004)}]{Casalbuoni:2003wh}%
  \BibitemOpen
  \bibfield  {author} {\bibinfo {author} {\bibfnamefont {R.}~\bibnamefont
  {Casalbuoni}}\ and\ \bibinfo {author} {\bibfnamefont {G.}~\bibnamefont
  {Nardulli}},\ }\href {\doibase 10.1103/RevModPhys.76.263} {\bibfield
  {journal} {\bibinfo  {journal} {Rev.Mod.Phys.}\ }\textbf {\bibinfo {volume}
  {76}},\ \bibinfo {pages} {263} (\bibinfo {year} {2004})},\ \Eprint
  {http://arxiv.org/abs/hep-ph/0305069} {arXiv:hep-ph/0305069 [hep-ph]}
  \BibitemShut {NoStop}%
\bibitem [{\citenamefont {Thies}(2006)}]{Thies:2006ti}%
  \BibitemOpen
  \bibfield  {author} {\bibinfo {author} {\bibfnamefont {M.}~\bibnamefont
  {Thies}},\ }\href {\doibase 10.1088/0305-4470/39/41/S04} {\bibfield
  {journal} {\bibinfo  {journal} {J.Phys.}\ }\textbf {\bibinfo {volume}
  {A39}},\ \bibinfo {pages} {12707} (\bibinfo {year} {2006})},\ \Eprint
  {http://arxiv.org/abs/hep-th/0601049} {arXiv:hep-th/0601049 [hep-th]}
  \BibitemShut {NoStop}%
\bibitem [{\citenamefont {Brzoska}\ and\ \citenamefont
  {Thies}(2002)}]{Brzoska:2001iq}%
  \BibitemOpen
  \bibfield  {author} {\bibinfo {author} {\bibfnamefont {A.}~\bibnamefont
  {Brzoska}}\ and\ \bibinfo {author} {\bibfnamefont {M.}~\bibnamefont
  {Thies}},\ }\href {\doibase 10.1103/PhysRevD.65.125001} {\bibfield  {journal}
  {\bibinfo  {journal} {Phys.Rev.}\ }\textbf {\bibinfo {volume} {D65}},\
  \bibinfo {pages} {125001} (\bibinfo {year} {2002})},\ \Eprint
  {http://arxiv.org/abs/hep-th/0112105} {arXiv:hep-th/0112105 [hep-th]}
  \BibitemShut {NoStop}%
\bibitem [{\citenamefont {Thies}(2004)}]{Thies:2003br}%
  \BibitemOpen
  \bibfield  {author} {\bibinfo {author} {\bibfnamefont {M.}~\bibnamefont
  {Thies}},\ }\href {\doibase 10.1103/PhysRevD.69.067703} {\bibfield  {journal}
  {\bibinfo  {journal} {Phys. Rev.}\ }\textbf {\bibinfo {volume} {D69}},\
  \bibinfo {pages} {067703} (\bibinfo {year} {2004})},\ \Eprint
  {http://arxiv.org/abs/hep-th/0308164} {arXiv:hep-th/0308164 [hep-th]}
  \BibitemShut {NoStop}%
\bibitem [{\citenamefont {Thies}\ and\ \citenamefont
  {Urlichs}(2003)}]{Thies:2003kk}%
  \BibitemOpen
  \bibfield  {author} {\bibinfo {author} {\bibfnamefont {M.}~\bibnamefont
  {Thies}}\ and\ \bibinfo {author} {\bibfnamefont {K.}~\bibnamefont
  {Urlichs}},\ }\href {\doibase 10.1103/PhysRevD.67.125015} {\bibfield
  {journal} {\bibinfo  {journal} {Phys. Rev.}\ }\textbf {\bibinfo {volume}
  {D67}},\ \bibinfo {pages} {125015} (\bibinfo {year} {2003})},\ \Eprint
  {http://arxiv.org/abs/hep-th/0302092} {arXiv:hep-th/0302092 [hep-th]}
  \BibitemShut {NoStop}%
\bibitem [{\citenamefont {Mermin}\ and\ \citenamefont {Wagner}(1966)}]{MW}%
  \BibitemOpen
  \bibfield  {author} {\bibinfo {author} {\bibfnamefont {N.~D.}\ \bibnamefont
  {Mermin}}\ and\ \bibinfo {author} {\bibfnamefont {H.}~\bibnamefont
  {Wagner}},\ }\href@noop {} {\bibfield  {journal} {\bibinfo  {journal} {Phys.
  Rev. Lett.}\ }\textbf {\bibinfo {volume} {17}},\ \bibinfo {pages} {1133}
  (\bibinfo {year} {1966})}\BibitemShut {NoStop}%
\bibitem [{\citenamefont {Hohenberg}(1967)}]{HB}%
  \BibitemOpen
  \bibfield  {author} {\bibinfo {author} {\bibfnamefont {P.~C.}\ \bibnamefont
  {Hohenberg}},\ }\href@noop {} {\bibfield  {journal} {\bibinfo  {journal}
  {Phys. Rev.}\ }\textbf {\bibinfo {volume} {158}},\ \bibinfo {pages} {383}
  (\bibinfo {year} {1967})}\BibitemShut {NoStop}%
\bibitem [{\citenamefont {Karbstein}\ and\ \citenamefont
  {Thies}(2007)}]{Karbstein:2006er}%
  \BibitemOpen
  \bibfield  {author} {\bibinfo {author} {\bibfnamefont {F.}~\bibnamefont
  {Karbstein}}\ and\ \bibinfo {author} {\bibfnamefont {M.}~\bibnamefont
  {Thies}},\ }\href {\doibase 10.1103/PhysRevD.75.025003} {\bibfield  {journal}
  {\bibinfo  {journal} {Phys. Rev.}\ }\textbf {\bibinfo {volume} {D75}},\
  \bibinfo {pages} {025003} (\bibinfo {year} {2007})},\ \Eprint
  {http://arxiv.org/abs/hep-th/0610243} {arXiv:hep-th/0610243 [hep-th]}
  \BibitemShut {NoStop}%
\bibitem [{\citenamefont {de~Forcrand}\ and\ \citenamefont
  {Philipsen}(2002)}]{deForcrand:2002ci}%
  \BibitemOpen
  \bibfield  {author} {\bibinfo {author} {\bibfnamefont {P.}~\bibnamefont
  {de~Forcrand}}\ and\ \bibinfo {author} {\bibfnamefont {O.}~\bibnamefont
  {Philipsen}},\ }\href {\doibase 10.1016/S0550-3213(02)00626-0} {\bibfield
  {journal} {\bibinfo  {journal} {Nucl.Phys.}\ }\textbf {\bibinfo {volume}
  {B642}},\ \bibinfo {pages} {290} (\bibinfo {year} {2002})},\ \Eprint
  {http://arxiv.org/abs/hep-lat/0205016} {arXiv:hep-lat/0205016 [hep-lat]}
  \BibitemShut {NoStop}%
\bibitem [{\citenamefont {Lombardo}(2006)}]{Lombardo:2006yc}%
  \BibitemOpen
  \bibfield  {author} {\bibinfo {author} {\bibfnamefont {M.}~\bibnamefont
  {Lombardo}},\ }\href@noop {} {\bibfield  {journal} {\bibinfo  {journal}
  {PoS}\ }\textbf {\bibinfo {volume} {CPOD2006}},\ \bibinfo {pages} {003}
  (\bibinfo {year} {2006})},\ \Eprint {http://arxiv.org/abs/hep-lat/0612017}
  {arXiv:hep-lat/0612017 [hep-lat]} \BibitemShut {NoStop}%
\bibitem [{\citenamefont {Philipsen}(2013)}]{Philipsen:2012nu}%
  \BibitemOpen
  \bibfield  {author} {\bibinfo {author} {\bibfnamefont {O.}~\bibnamefont
  {Philipsen}},\ }\href {\doibase 10.1016/j.ppnp.2012.09.003} {\bibfield
  {journal} {\bibinfo  {journal} {Prog.Part.Nucl.Phys.}\ }\textbf {\bibinfo
  {volume} {70}},\ \bibinfo {pages} {55} (\bibinfo {year} {2013})},\ \Eprint
  {http://arxiv.org/abs/1207.5999} {arXiv:1207.5999 [hep-lat]} \BibitemShut
  {NoStop}%
\bibitem [{\citenamefont {Ebert}\ \emph {et~al.}(2011)\citenamefont {Ebert},
  \citenamefont {Gubina}, \citenamefont {Klimenko}, \citenamefont {Kurbanov},\
  and\ \citenamefont {Zhukovsky}}]{Ebert:2011rg}%
  \BibitemOpen
  \bibfield  {author} {\bibinfo {author} {\bibfnamefont {D.}~\bibnamefont
  {Ebert}}, \bibinfo {author} {\bibfnamefont {N.}~\bibnamefont {Gubina}},
  \bibinfo {author} {\bibfnamefont {K.}~\bibnamefont {Klimenko}}, \bibinfo
  {author} {\bibfnamefont {S.}~\bibnamefont {Kurbanov}}, \ and\ \bibinfo
  {author} {\bibfnamefont {V.~C.}\ \bibnamefont {Zhukovsky}},\ }\href {\doibase
  10.1103/PhysRevD.84.025004} {\bibfield  {journal} {\bibinfo  {journal} {Phys.
  Rev.}\ }\textbf {\bibinfo {volume} {D84}},\ \bibinfo {pages} {025004}
  (\bibinfo {year} {2011})},\ \Eprint {http://arxiv.org/abs/1102.4079}
  {arXiv:1102.4079 [hep-ph]} \BibitemShut {NoStop}%
\bibitem [{\citenamefont {Gubina}\ \emph {et~al.}(2012)\citenamefont {Gubina},
  \citenamefont {Klimenko}, \citenamefont {Kurbanov},\ and\ \citenamefont
  {Zhukovsky}}]{Gubina:2012wp}%
  \BibitemOpen
  \bibfield  {author} {\bibinfo {author} {\bibfnamefont {N.}~\bibnamefont
  {Gubina}}, \bibinfo {author} {\bibfnamefont {K.}~\bibnamefont {Klimenko}},
  \bibinfo {author} {\bibfnamefont {S.}~\bibnamefont {Kurbanov}}, \ and\
  \bibinfo {author} {\bibfnamefont {V.~C.}\ \bibnamefont {Zhukovsky}},\ }\href
  {\doibase 10.1103/PhysRevD.86.085011} {\bibfield  {journal} {\bibinfo
  {journal} {Phys. Rev.}\ }\textbf {\bibinfo {volume} {D86}},\ \bibinfo {pages}
  {085011} (\bibinfo {year} {2012})},\ \Eprint {http://arxiv.org/abs/1206.2519}
  {arXiv:1206.2519 [hep-ph]} \BibitemShut {NoStop}%
\bibitem [{\citenamefont {Ebert}\ \emph {et~al.}(2014)\citenamefont {Ebert},
  \citenamefont {Khunjua}, \citenamefont {Klimenko},\ and\ \citenamefont
  {Zhukovsky}}]{Ebert:2013dda}%
  \BibitemOpen
  \bibfield  {author} {\bibinfo {author} {\bibfnamefont {D.}~\bibnamefont
  {Ebert}}, \bibinfo {author} {\bibfnamefont {T.}~\bibnamefont {Khunjua}},
  \bibinfo {author} {\bibfnamefont {K.}~\bibnamefont {Klimenko}}, \ and\
  \bibinfo {author} {\bibfnamefont {V.~C.}\ \bibnamefont {Zhukovsky}},\ }\href
  {\doibase 10.1142/S0217751X14500250} {\bibfield  {journal} {\bibinfo
  {journal} {Int. J. Mod. Phys. Rev.Phys.}\ }\textbf {\bibinfo {volume}
  {A29}},\ \bibinfo {pages} {1450025} (\bibinfo {year} {2014})},\ \Eprint
  {http://arxiv.org/abs/1306.4485} {arXiv:1306.4485 [hep-th]} \BibitemShut
  {NoStop}%
\bibitem [{\citenamefont {Feinberg}(2004)}]{Feinberg:2003qz}%
  \BibitemOpen
  \bibfield  {author} {\bibinfo {author} {\bibfnamefont {J.}~\bibnamefont
  {Feinberg}},\ }\href {\doibase 10.1016/j.aop.2003.08.004} {\bibfield
  {journal} {\bibinfo  {journal} {Annals Phys.}\ }\textbf {\bibinfo {volume}
  {309}},\ \bibinfo {pages} {166} (\bibinfo {year} {2004})},\ \Eprint
  {http://arxiv.org/abs/hep-th/0305240} {arXiv:hep-th/0305240 [hep-th]}
  \BibitemShut {NoStop}%
\bibitem [{\citenamefont {Gross}\ and\ \citenamefont
  {Neveu}(1974)}]{Gross:1974jv}%
  \BibitemOpen
  \bibfield  {author} {\bibinfo {author} {\bibfnamefont {D.~J.}\ \bibnamefont
  {Gross}}\ and\ \bibinfo {author} {\bibfnamefont {A.}~\bibnamefont {Neveu}},\
  }\href {\doibase 10.1103/PhysRevD.10.3235} {\bibfield  {journal} {\bibinfo
  {journal} {Phys.Rev.}\ }\textbf {\bibinfo {volume} {D10}},\ \bibinfo {pages}
  {3235} (\bibinfo {year} {1974})}\BibitemShut {NoStop}%
\bibitem [{\citenamefont {Hubbard}(1959)}]{Hubbard:1959ub}%
  \BibitemOpen
  \bibfield  {author} {\bibinfo {author} {\bibfnamefont {J.}~\bibnamefont
  {Hubbard}},\ }\href {\doibase 10.1103/PhysRevLett.3.77} {\bibfield  {journal}
  {\bibinfo  {journal} {Phys. Rev. Lett.}\ }\textbf {\bibinfo {volume} {3}},\
  \bibinfo {pages} {77} (\bibinfo {year} {1959})}\BibitemShut {NoStop}%
\bibitem [{\citenamefont {Stratonovich}(1957)}]{Stratonovich}%
  \BibitemOpen
  \bibfield  {author} {\bibinfo {author} {\bibfnamefont {R.}~\bibnamefont
  {Stratonovich}},\ }\href@noop {} {\bibfield  {journal} {\bibinfo  {journal}
  {Dokl. Akad. Nauk.}\ }\textbf {\bibinfo {volume} {115}},\ \bibinfo {pages}
  {1097} (\bibinfo {year} {1957})}\BibitemShut {NoStop}%
\bibitem [{\citenamefont {Pausch}\ \emph {et~al.}(1991)\citenamefont {Pausch},
  \citenamefont {Thies},\ and\ \citenamefont {Dolman}}]{Pausch:1991ee}%
  \BibitemOpen
  \bibfield  {author} {\bibinfo {author} {\bibfnamefont {R.}~\bibnamefont
  {Pausch}}, \bibinfo {author} {\bibfnamefont {M.}~\bibnamefont {Thies}}, \
  and\ \bibinfo {author} {\bibfnamefont {V.}~\bibnamefont {Dolman}},\ }\href
  {\doibase 10.1007/BF01295773} {\bibfield  {journal} {\bibinfo  {journal}
  {Z.Phys.}\ }\textbf {\bibinfo {volume} {A338}},\ \bibinfo {pages} {441}
  (\bibinfo {year} {1991})}\BibitemShut {NoStop}%
\bibitem [{\citenamefont {Wolff}(1985)}]{Wolff:1985av}%
  \BibitemOpen
  \bibfield  {author} {\bibinfo {author} {\bibfnamefont {U.}~\bibnamefont
  {Wolff}},\ }\href {\doibase 10.1016/0370-2693(85)90671-9} {\bibfield
  {journal} {\bibinfo  {journal} {Phys.Lett.}\ }\textbf {\bibinfo {volume}
  {B157}},\ \bibinfo {pages} {303} (\bibinfo {year} {1985})}\BibitemShut
  {NoStop}%
\bibitem [{\citenamefont {Nambu}\ and\ \citenamefont
  {Jona-Lasinio}(1961{\natexlab{a}})}]{Nambu:1961tp}%
  \BibitemOpen
  \bibfield  {author} {\bibinfo {author} {\bibfnamefont {Y.}~\bibnamefont
  {Nambu}}\ and\ \bibinfo {author} {\bibfnamefont {G.}~\bibnamefont
  {Jona-Lasinio}},\ }\href {\doibase 10.1103/PhysRev.122.345} {\bibfield
  {journal} {\bibinfo  {journal} {Phys.Rev.}\ }\textbf {\bibinfo {volume}
  {122}},\ \bibinfo {pages} {345} (\bibinfo {year}
  {1961}{\natexlab{a}})}\BibitemShut {NoStop}%
\bibitem [{\citenamefont {Nambu}\ and\ \citenamefont
  {Jona-Lasinio}(1961{\natexlab{b}})}]{Nambu:1961fr}%
  \BibitemOpen
  \bibfield  {author} {\bibinfo {author} {\bibfnamefont {Y.}~\bibnamefont
  {Nambu}}\ and\ \bibinfo {author} {\bibfnamefont {G.}~\bibnamefont
  {Jona-Lasinio}},\ }\href {\doibase 10.1103/PhysRev.124.246} {\bibfield
  {journal} {\bibinfo  {journal} {Phys.Rev.}\ }\textbf {\bibinfo {volume}
  {124}},\ \bibinfo {pages} {246} (\bibinfo {year}
  {1961}{\natexlab{b}})}\BibitemShut {NoStop}%
\end{thebibliography}%

\end{document}